\documentclass[aps,pre,twocolumn,showpacs,floatfix]{revtex4} 
\usepackage{amsmath,amssymb,isolatin1,graphicx,times}

\newcommand{\be}{\begin{equation}}
\newcommand{\ee}{\end{equation}}
\newcommand{\bea}{\begin{eqnarray}}
\newcommand{\eea}{\end{eqnarray}}
\newcommand{\bw}{\begin{widetext}}
\newcommand{\ew}{\end{widetext}}

\newcommand{\pcl}{{\cal P}_{\rm cl}}
\newcommand{\pqm}{{\cal P}_{\rm qm}}
\newcommand{\delp}{\Delta{\cal P}}
\newcommand{\maxa}{{\rm env}[|\overline{\alpha}(t)|^2]}
\newcommand{\maxad}{{\rm env}[|\overline{\alpha}_{\rm discr}(t)|^2]}
\newcommand{\maxp}{{\rm env}[\overline{\pi}](t)}

\newcommand{\kommentar}[1]{}

\begin{document}
 
\title{Efficiency of quantum and classical transport on graphs}
\author{Oliver M{\"u}lken}
\email{oliver.muelken@physik.uni-freiburg.de}
\author{Alexander Blumen}
\affiliation{
Theoretische Polymerphysik, Universit\"at Freiburg,
Hermann-Herder-Straße 3, 79104 Freiburg i.Br., Germany}

\date{\today} 
\begin{abstract}
We propose a measure to quantify the efficiency of classical and quantum
mechanical transport processes on graphs. The measure only depends on the
density of states (DOS), which contains all the necessary information
about the graph. For some given (continuous) DOS, the measure shows a
power law behavior, where the exponent for the quantum transport is twice
the exponent of its classical counterpart. For small-world networks,
however, the measure shows rather a stretched exponential law but still
the quantum transport outperforms the classical one. Some finite
tree-graphs have a few highly degenerate eigenvalues, such that, on the
other hand, on them the classical transport may be more efficient than the
quantum one.
\end{abstract}
\maketitle

\section{Introduction.}
The transfer of information is the cornerstone of many physical, chemical
or biological processes. The information can be encoded in the mass,
charge or energy transported.  All these transfer processes depend on the
underlying structure of the system under study. These could be, for
example, simple crystals, as in solid state physics \cite{Ziman}, more
complex molecular aggregates like polymers \cite{Kenkre}, or general
network structures \cite{albert2002}. Of course, there exists a panoply of
further chemical or biological systems which propagate information.

There are several approaches to model the transport on these structures.
In (quantum) mechanics, the structure, i.e., the potential a particle is
moving in, specifies the Hamiltonian of the system, which determines the
time evolution. For instance, the dynamics of an electron in a simple
crystal is described by the Bloch ansatz \cite{Ziman}. H{\"u}ckel's
molecular-orbital theory in quantum chemistry allows to define a
Hamiltonian for more complex structures, such as molecules
\cite{McQuarrie}. This is again related to transport processes in
polymers, where the connectivity of the polymer plays a fundamental role
in its dynamical and relaxational properties \cite{Doi-Edwards}. There,
(classical) transport processes can be described by a master equation
approach with an appropriate (classical) transfer operator which
determines the temporal evolution of an excitation \cite{Kenkre,weiss}.  

In all examples listed above, the densities of states (DOS), or spectral
density, of a given system of size $N$, 
\[
\rho(\lambda) =
\frac{1}{N}\sum_{n=1}^N \delta(\lambda-\lambda_n),
\]
contains the essential
informations about the system. Here, the $\lambda_n$'s are the eigenvalues
of the appropriate Hamiltonian ${\bf H}$ or transfer operator ${\bf T}$.
Depending (mainly) on the topology of the system, $\rho(\lambda)$ shows
very distinct features.  A classic in this respect is the DOS of a random
matrix, corresponding to a random graph \cite{Mehta}. Wigner has shown
that for a (large) matrix with (specific) random entries, the eigenvalues
of this matrix lie within a semi-circle \cite{wigner1955}.  As we will
show, distinct features of the DOS also result in very distinct transport
properties.

\section{Transport on graphs.} 
We start our discussion by considering quantum mechanical transport
processes on discrete structures, in general called graphs, which are a
collection of $N$ connected nodes. We assume that the states $|j\rangle$,
associated with a localized excitation at node $j$, form an orthonormal
basis set and span the whole accessible Hilbert space. The time evolution
of an excitation initially placed at node $|j\rangle$ is determined by the
systems' Hamiltonian ${\bf H}$ and reads $\exp(-i{\bf H}t)|j\rangle$.  The
classical transport can be described by a master equation for the
conditional probability, $p_{k,j}(t)$, to find an excitation at time $t$
at node $k$ when starting at time $0$ at node $j$. Using also here the
Dirac notation for a state at node $j$, the classical time evolution of
this state follows from the transfer matrix ${\bf T}$ of the transport
process as $\exp({\bf T}t)|j\rangle$. In order to compare the classical
and the quantum motion, we identify the Hamiltonian of the system with the
(classical) transfer matrix, ${\bf H} = - {\bf T}$, which we will relate
later to the (discrete) Laplacian of the graph, see e.g.\
\cite{farhi1998,mb2005a}.  The classical and quantum mechanical transition
probabilities to go from the state $|j\rangle$ at time $0$ to the state
$|k\rangle$ in time $t$ are given by $ p_{k,j}(t) \equiv \langle k |
\exp({\bf T} t) | j \rangle $ and $ \pi_{k,j}(t) \equiv
|\alpha_{k,j}(t)|^2 \equiv |\langle k | \exp(- i {\bf H} t) | j \rangle|^2
$, respectively.

\section{Averaged transition probabilities.}
Quantum mechanically, a lower bound of the {\sl average} probability to be
still or again at the initially excited node, $\overline{\pi}_{\rm
discr}(t) \equiv \frac{1}{N} \sum_{j=1}^{N} \pi_{j,j}(t)$, is obtained for
a finite network by an eigenstate expansion and using the Cauchy-Schwarz
inequality as, \cite{mbb2006a},
\be
\overline{\pi}_{\rm discr}(t) \geq
\left|\frac{1}{N} \sum_{n} \ \exp(-i\lambda_nt) \right|^2 \equiv
|\overline{\alpha}_{\rm discr}(t)|^2.
\label{pqmavg}
\ee
Note that $|\overline{\alpha}_{\rm discr}(t)|^2$ depends {\sl only} on the
eigenvalues of ${\bf H}$ but {\sl not} on the eigenvectors. As we have
shown earlier, especially the local maxima of $\overline{\pi}(t)$ are very
well reproduced by $|\overline{\alpha}(t)|^2$ and for regular networks,
the lower bound is exact \cite{mbb2006a}. Therefore, we will use
$|\overline{\alpha}_{\rm discr}(t)|^2$ in the following to characterize
transport processes.

Also classically one has a simple expression for $\overline{p}_{\rm
discr}(t) \equiv \frac{1}{N} \sum_{j=1}^{N} p_{j,j}(t)$, see, e.g.,
\cite{bray1988},
\be
\overline{p}_{\rm discr}(t) =
\frac{1}{N} \sum_{n=1}^{N} \ \exp\big(-\lambda_n t\big).
\label{pclavg}
\ee 
Again, this result depends only on the (discrete) eigenvalue spectrum of
${\bf T}$ but {\sl not} on the eigenvectors.

In the continuum limit, Eqs.~(\ref{pqmavg}) and (\ref{pclavg}) can be
written as
\bea
\overline{\pi}(t) &\ge& \left| \int d\lambda \ \rho(\lambda) \
\exp(-i \lambda t) \right|^2 \equiv |\overline{\alpha}(t)|^2
.
\label{pqmavginf} \\
\overline{p}(t) &=& \int d\lambda \ \rho(\lambda) \
\exp(-\lambda t), 
\label{pclavginf} 
\eea
The explicit calculation of the integrals is easily done using computer
algebra systems like MAPLE or MATHEMATICA; in many cases, the integrals can
also be found in \cite{gradshteyn}.

\section{Efficiency measure of transport on graphs.} 
Equations (\ref{pqmavg})-(\ref{pclavginf}) allow to define an efficiency
measure (EM) for the performance of the transport on a graph.  We stress
again, that the EM does not involve any computationally expensive
calculations of eigenstates. Rather, only the energy eigenvalues are
needed, which are quite readily obtained by diagonalizing {\bf H}.

By starting with continuous DOS, since those are mathematically easier to
handle, we define the (classical) EM of the graph by the decay of
$\overline{p}(t)$ for large $t$, where a fast decay means that the initial
excitation spreads rapidly over the whole graph. Quantum mechanically,
however, the transition probabilities fluctuate due to the unitary time
evolution.  Therefore, in most cases also $\overline{\pi}(t)$ and
$|\overline{\alpha}(t)|^2$ fluctuate.  Nevertheless, the local maxima of
$|\overline{\alpha}(t)|^2$ reproduce the ones of $\overline{\pi}(t)$
rather well.  We use now the temporal scaling of the local maxima of
$|\overline{\alpha}(t)|^2$ as the (quantum) EM and denote the envelope of
the maxima by $\maxa$.  Similar to the classical case, a fast decay of
$\maxa$ corresponds to a rapid spreading of an initial excitation. 

For a large variety of graphs the DOS can be written as 
\be
\rho(\lambda)
\sim (\lambda\lambda_m-\lambda^2)^\nu,
\ee
with $\nu>-1$ and where
$\lambda_m$ is the maximal eigenvalue (we assumed the minimum eigenvalue
to be zero). Since we are interested in the large $t$ behavior,
$\overline{p}(t)$ [and also $\overline{p}_{\rm discr}(t)$] will be mainly
determined by small $\lambda$ values, such that for $t\gg1$ we can assume
$\rho(\lambda) \sim \lambda^\nu$. Than it is easy to show that the
classical EM scales as
\be
\overline{p}(t)\sim t^{-(1+\nu)}.
\label{pl_cl}
\ee
This scaling argument for long times is well known throughout the
literature, where $2(1+\nu)\equiv d_s$ is sometimes called the spectral or
fracton dimension, see, e.g., \cite{alexander1981}. 

In order to obtain the quantum mechanical scaling for the same DOS, we can
use the same scaling arguments. For $t\gg1$, also
$|\overline{\alpha}(t)|^2$ [and $|\overline{\alpha}_{\rm discr}(t)|^2$]
will be mainly determined by the small $\lambda$ values. In fact, for
$\rho(\lambda) \sim \lambda^\nu$ one has
$|\overline{\alpha}(t)|=\overline{p}(t)$. Here, all quantum mechanical
oscillations vanish, because we consider only the leading term of the DOS
for small $\lambda$. Thus, we furthermore have $\maxa =
|\overline{\alpha}(t)|^2$, i.e., the quantum EM reads 
\be
\maxa \sim t^{-2(1+\nu)}.
\label{pl_qm}
\ee
Equation (\ref{pl_qm}) can also be directly obtained from Eq.\
(\ref{pqmavginf}) with $\rho(\lambda) \sim \lambda^\nu$.  Of course, Eqs.\
(\ref{pl_cl}) and (\ref{pl_qm}) agree with the solution for
$\overline{p}(t)$ and $\maxa$ obtained from the full DOS $\rho(\lambda)
\sim (\lambda\lambda_m-\lambda^2)^\nu$.  The same scaling has been
obtained for the decay of temporal correlations in quantum mechanical
systems with Cantor spectra \cite{ketzmerick92}.  There, the (full)
probability $\overline{\pi}(t)$, which was smoothed over time, was used.

In general, for $\overline{p}(t) \sim t^{-\pcl}$, the exponent $\pcl$
determines the classical EM of the graph because larger $\pcl$
correspond to a faster decay of $\overline{p}(t)$.  Quantum mechanically,
we may have $\maxa \sim t^{-\pqm}$, such that the exponent $\pqm$
determines the quantum EM of the graph. Since we consider only
the local maxima, the actual (fluctuating) probability $\overline{\pi}(t)$
[bounded from below by $|\overline{\alpha}(t)|^2$] might drop well below
these values, i.e., there are times $t$ at which $\overline{\pi}(t) \ll
1$. However, these values are very localized in time and the overall
performance of the quantum transport is best quantified by the scaling of
$\maxa$.

The difference between the classical and quantum EM is given by
the factor 
\be
\delp(t) \equiv \ln [\maxa] / \ln [\overline{p}(t)].
\ee
For
classical and quantum power law behavior $\delp(t)$ is time-independent
and we have $\delp = \pqm/\pcl$. Thus, for the DOS given above, with $\nu
< \infty$, we get $\delp=2$, as could be expected from the wave-like
behavior of the quantum motion compared to the normal diffusive behavior of
the classical motion.  

\subsection{Continuous DOS.}
Two important examples are connected to scaling. An infinite hypercubic
lattice in $d$ dimensions has as eigenvalues
$\lambda(\Theta_1,\dots,\Theta_d) \equiv \sum_{n=1}^d \lambda(\Theta_n)$,
with $\lambda(\Theta_n) = 2- 2\cos\Theta_n$ and $\Theta_n\in[0,2\pi[$.
Here, one can calculate explicitly $|\overline{\alpha}(t)|^2$ and
$\overline{\pi}(t)$ and demonstrate that the local maxima really obey
scaling; we get namely
$|\overline{\alpha}(t)|^2=\overline{\pi}(t)\sim|J_0(2t)|^{2d}$
\cite{mbb2006a}.  For $t\gg1$ this can be approximated by
$\overline{\pi}(t)\sim \sin^{2d}(2t+\pi/4)/t^d$ \cite{gradshteyn}. Since
the maximum of the $\sin$-function is $1$, the quantum measure scales as
$\maxa = \maxp \sim t^{-d}$, which is what one also obtains from the
scaling argument above and $\rho(\lambda)\sim\lambda^{d/2-1}$.  Then
$\nu=d/2-1$, and the classical measure scales as $\overline{p}(t) \sim
t^{-d/2}$ (i.e., the spectral dimension is $d_s=d$). 

As a second example, we take a random graph. It was shown that the
eigenvalue spectrum of the Laplacian of such a graph obeys Wigner's
semi-circle law \cite{wigner1955,Mehta}, which we obtain for $\nu=1/2$
from the DOS given above. For large times both measures again obey scaling
and we have $\overline{p}(t)\sim t^{-3/2}$ and $\maxa\sim t^{-3}$. 

\begin{figure}[htb]
\centerline{\includegraphics[clip=,width=\columnwidth]{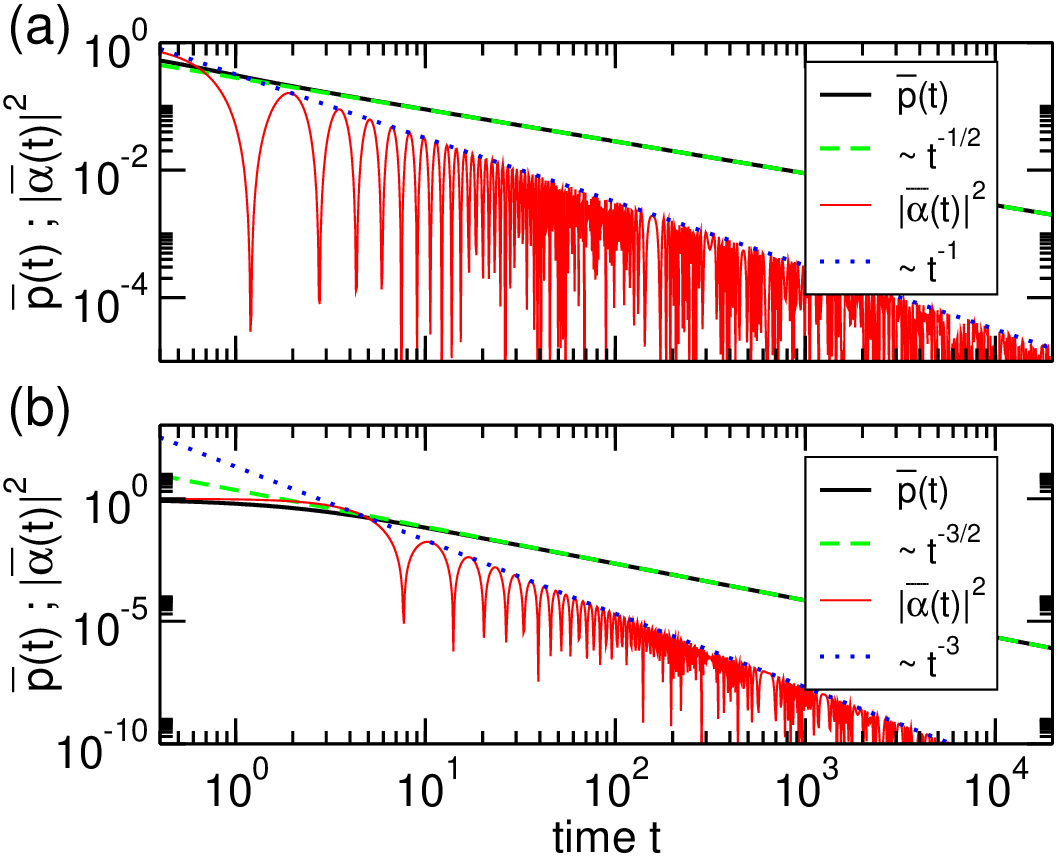}} 
\caption{(Color online). $\overline{p}(t)$ and $|\overline{\alpha}(t)|^2$
as well as the power laws given in Eqs.~(\ref{pl_qm}) and (\ref{pl_cl})
for (a) an infinite regular (1D) graph ($\nu=-1/2$) and (b) a random graph
whose DOS obeys Wigner's semi-circle law ($\nu=1/2$).}
\label{percolability}
\end{figure}

Figure \ref{percolability} shows the temporal behavior of
$\overline{p}(t)$ and $|\overline{\alpha}(t)|^2$ as well as the power law
behavior of $\overline{p}(t)$ and $\maxa$ for (a) an infinite, regular,
one-dimensional (1D) graph and (b) a random graph. Note that here (and in
the following figures, too) the very localized minima of
$|\overline{\alpha}(t)|^2$ do not always show up clearly in the
logarithmic scale used.

For some DOS, the EMs show no power law behavior.  The DOS given above are
bounded from above by a maximal eigenvalue. This does not have to be the
case. The DOS of small-world networks, for instance, may show long
$\lambda$-tails \cite{monasson1999}.  One additional feature of such DOS
is that they do not obey any simple scaling for small $\lambda$.
Nevertheless, sometimes analytic solutions for, at least,
$\overline{p}(t)$ can be obtained \cite{monasson1999}, as, for example for
certain 1D systems with
$\rho(\lambda)\sim\lambda^{-3/2}\exp(-1/\sqrt{\lambda})$.

For computational simplicity we consider a 2D system with 
\be
\rho(\lambda) =
\lambda^{-b}\exp(-1/\lambda)
\ee
for $\lambda \in [0,\infty[$ and $b>1$. The
term $\exp(-1/\lambda)$ is usually referred to as Lifshits tail, while the
term $\lambda^{-b}$ assures that $\lim_{\lambda\to\infty}\rho(\lambda)=0$.
Then, for $t\gg1$, the EMs are proportional to the product of a stretched
exponential and a power law \cite{gradshteyn}, 
\bea
|\overline{\alpha}(t)|^2 = \maxa &\sim&
t^{(2b-3)/2}\exp\left(-2\sqrt{2t}\right)
\label{pl_qm_dis}\\
\overline{p}(t) &\sim& t^{(2b-3)/4}\exp\left(-2\sqrt{t}\right) 
\label{pl_cl_dis}.
\eea
Furthermore, $\overline{\pi}(t)$ does {\sl not} oscillate, an effect which
is interesting in itself but we will not elaborate on this here. 

Although we do not obtain a simple relation between the classical and the
quantum EMs, $[\overline{p}(t)]^2$ and $\maxa$ still display similar
functional forms. Now, however, $\delp(t) = [2(2b-3)\ln t
- 8\sqrt{2t}]/[(2b-3)\ln t - 8\sqrt{t}]$ is time-dependent.  Equations
  (\ref{pl_cl_dis}) and (\ref{pl_qm_dis}) are only valid for $t\gg1$, such
that $\lim_{t\to\infty} \delp(t) = \sqrt{2}$ for all $b$. Hence, also here
the quantum transport outperforms the classical one, which is also
confirmed by numerical integration of the time-dependent Schr\"odinger
equation for a small-world network \cite{kim2003}.  In fact, in both cases
the transport is faster than for a regular 2D graph. We note that
localization is related to other features of the DOS as we will recall
below.  However, at intermediate times the quantum EM may drop below the
classical EM; the position of the crossover from $\delp(t) < 1$ to
$\delp(t) >1$ depends on the exponent $b$. 

\subsection{Discrete DOS.}
Up to now, we have only considered continuous DOS, where the quantum EM is
quicker than the classical one. In the following we will consider discrete
DOS which are obtained by modeling the motion on a given graph classically
by continuous-time random walks (CTRWs), see, e.g., \cite{weiss}, and
quantum mechanically by continuous-time quantum walks (CTQWs)
\cite{farhi1998,mb2005a}. The Hamiltonian is given by the (discrete)
Laplacian associated with the graph, i.e., by the functionality of the
nodes and their connectivity. We assume the jump rates between all
connected pairs of nodes of the graph to be equal.

\begin{figure}[htb]
\centerline{\includegraphics[clip=,width=\columnwidth]{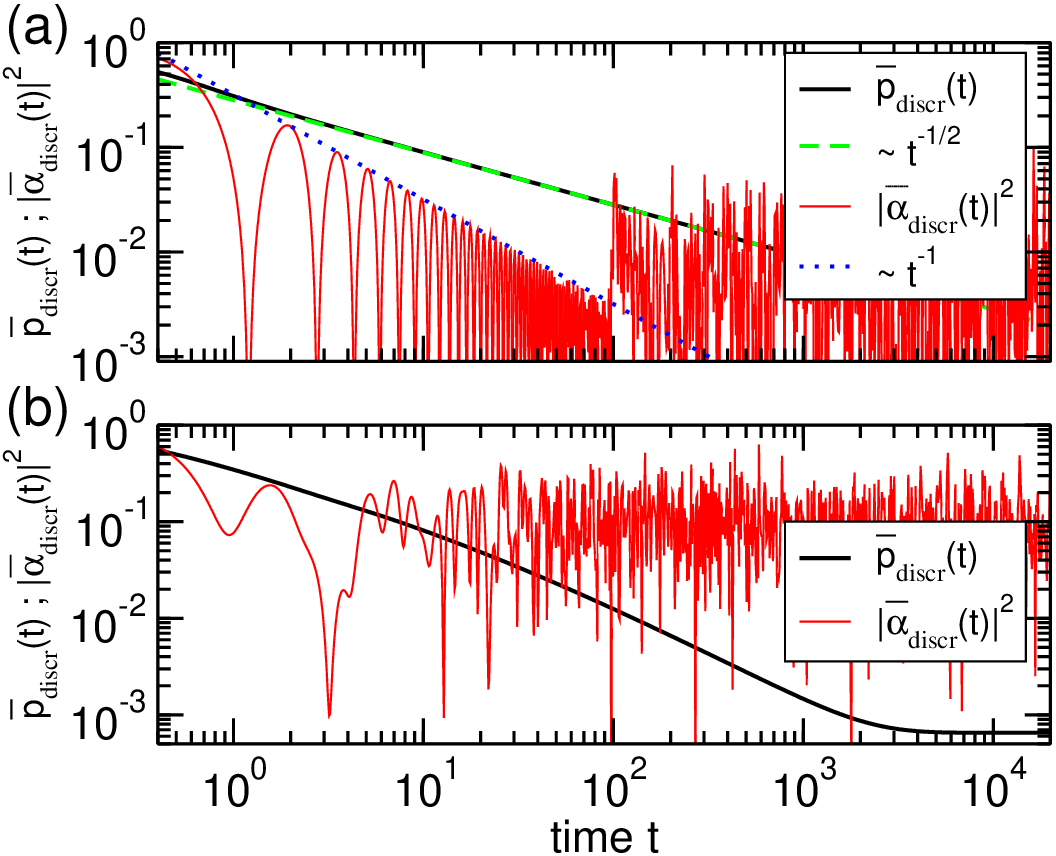}} 
\caption{(Color online). $\overline{p}_{\rm discr}(t)$ and
$|\overline{\alpha}_{\rm discr}(t)|^2$ for (a) a finite regular (1D) graph
of size $N=200$ with periodic boundary conditions and (b) a dendrimer of
generation $10$ having functionality $z=3$, i.e.\ $N=3\cdot 2^{10} -2$.
Panel (a) contains also the power-law behavior for the infinite regular
(1D) graph, see Fig.~\ref{percolability}(a).}
\label{percolability_finite}
\end{figure}

In general, for finite graphs, $\overline{p}_{\rm discr}(t)$ and $\maxad$
do not decay ad infinitum but at some time will remain constant
(classically) or fluctuate about a constant value (quantum mechanically).
This time is given by the time it takes for the CTRW to reach the
(equilibrium) equipartitioned probability distribution and for the CTQW to
fluctuate about a saturation value. At intermediate times,
$\overline{p}_{\rm discr}(t)$ and $\maxad$ will show the same scaling as
for a system with the corresponding continuous DOS. Figure
\ref{percolability_finite}(a) shows the temporal behavior of
$\overline{p}_{\rm discr}(t)$ and $|\overline{\alpha}_{\rm discr}(t)|^2$
for a finite regular 1D graph of size $N=200$ with periodic boundary
conditions, see also \cite{mb2005b}. At intermediate times, the scaling
behavior is obviously that of the continuous case shown in Fig.\
\ref{percolability}(a).

Tree-like graphs do not display scaling in general. For CTQW on
hyperbranched structures (like Cayley trees, dendrimers or Husimi cacti),
the transition probability between two nodes strongly depends on the site
$j$ of the initial excitation \cite{mb2005a,mbb2006a}. Even in the long
time average, 
\be
\chi_{k,j} = \lim_{T\to\infty} T^{-1} \int_0^T dt \
\pi_{k,j}(t),
\ee
there are transition probabilities which are considerably
lower than the equipartitioned classical value \cite{mb2005a,mbb2006a}.
In Fig.\ \ref{percolability_finite}(b) we display the temporal behavior of
$\overline{p}_{\rm discr}(t)$ and $|\overline{\alpha}_{\rm discr}(t)|^2$
for a dendrimer of generation $10$ having functionality $z=3$, i.e.\
$N=3\cdot 2^{10} -2$. Here the classical curve does not show scaling at
intermediate times. Quantum mechanically, however,
$|\overline{\alpha}_{\rm discr}(t)|^2$ has a strong dip at short times but
then fluctuates about a finite value which is larger than the classical
saturation value. One should also bear in mind that
$|\overline{\alpha}_{\rm discr}(t)|^2$ is a lower bound and the actual
probability will be larger. Therefore, according to our measure for
intermediate $t\gg1$, the classical transport outperforms the
quantum transport on these special, {\sl finite} graphs. As we proceed to
show, the reason for this is to be found in the DOS.  This is related to
(Anderson) localization. Anderson showed that for localization the DOS has
to display a discrete finite series of $\delta$ functions
\cite{anderson1978}.

We consider now a simple star graph, having one core node and $N-1$ nodes
directly connected to the core but not to each other.  The eigenvalue
spectrum of this star has a very simple structure, there are $3$ distinct
eigenvalues, namely $\lambda_1=0$, $\lambda_2=1$, and $\lambda_3=N$,
having as degeneracies $g_1=1$, $g_2=N-2$, and $g_3=1$.  Therefore we get
\bea
\overline{p}_{\rm discr}(t) &=& \frac{1}{N} \left[ 1 + (N-2) e^{-t} +
e^{-(N-2)t} \right] \\
\overline{\pi}_{\rm discr}(t) &\ge& \frac{1}{N^2} \left| 1 + (N-2) e^{-it}
+ e^{-i(N-2)t} \right|^2.
\label{pi_star}
\eea
Obviously, only the term $|(N-2)\exp(-it)|^2/N^2 = (N-2)^2/N^2$ in Eq.\
(\ref{pi_star}) is of order ${\cal O}(1)$.  All the other terms are of
order ${\cal O}(1/N)$ or ${\cal O}(1/N^2)$ and, therefore, cause only
small oscillations (fluctuating terms) about or negligible shifts
(constant terms) from $(N-2)^2/N^2$. 

\begin{figure}[htb]
\centerline{\includegraphics[clip=,width=\columnwidth]{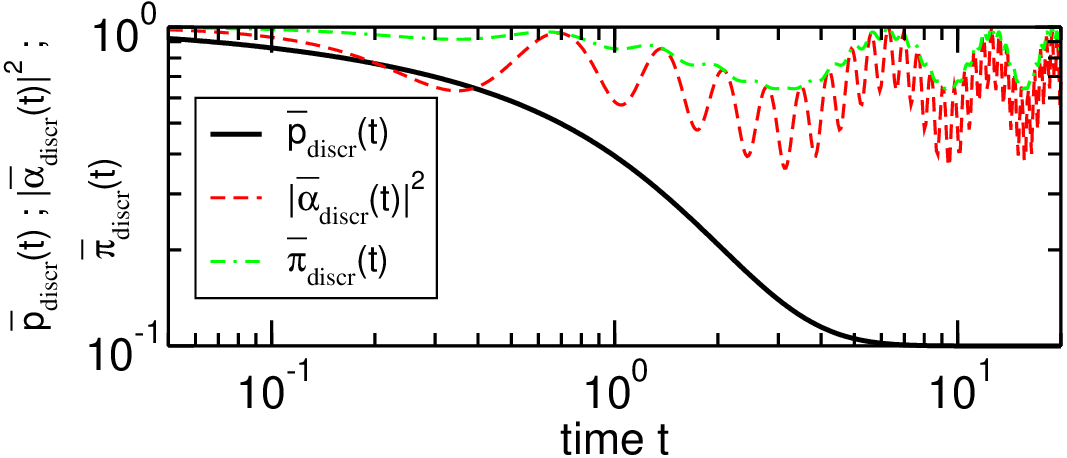}} 
\caption{(Color online). $\overline{p}_{\rm discr}(t)$,
$\overline{\pi}_{\rm discr}(t)$, and $|\overline{\alpha}_{\rm
discr}(t)|^2$ for a star with $N=10$.
}
\label{percolability_star}
\end{figure}

Having only {\sl one} low lying eigenvalue which is highly degenerate and
no other eigenvalue of a degeneracy of the same order of magnitude,
results in $\overline{p}_{\rm discr}(t) < \overline{\pi}_{\rm discr}(t)$
for all times $t$ and $\overline{p}_{\rm discr}(t) <
|\overline{\alpha}_{\rm discr}(t)|^2$ for almost all times $t$. Figure
\ref{percolability_star} shows the temporal behavior of $\overline{p}_{\rm
discr}(t)$, $\overline{\pi}_{\rm discr}(t)$, and $|\overline{\alpha}_{\rm
discr}(t)|^2$ for $N=10$. Now, for all times, the quantum transport is
slower than the classical one. We also see that $|\overline{\alpha}_{\rm
discr}(t)|^2$ fluctuates about $(N-2)^2/N^2 = 16/25$.

In general we find for our star-graph that the classical EM is lower than
the quantum EM. This result is to some extent also obeyed by dendrimers
and by other hyperbranched structures. These, too, have a few highly
degenerate eigenvalues, all other degeneracies being an order of magnitude
less, which results in the absence of any scaling of $\maxad$, see Fig.\
\ref{percolability_finite}(b). Of course, the details are much more
complex due to the more complex structure, we will elaborate on this
elsewhere. 

\section{Conclusion.}
We have proposed a measure to classify the efficiency of classical and
quantum mechanical transport processes. Depending on the density of
states, the quantum transport outperforms the classical transport by means
of the speed of the spreding of an initial excitation over a given system.
For algebraic DOS, the EMs confirm the difference between classical
diffusive and quantum mechanical wave-like transport.  Also for
small-world networks the quantum mechanical EM is lower than the classical
one, i.e.\ the quantum mechanical transport is faster.

However, for some finite graphs with a few highly degenerate eigenvalues it
may happen that the classical transport is more efficient, i.e., that the
(quantum) states become localized. We have shown this analytically for a
simple star graph. More complex structures, like dendrimers or
hyperbranched fractals, show an analogous behavior. 

\section*{Acknowledgments.}
We thank Veronika Bierbaum for producing the data for Fig.\
\ref{percolability_finite}(b).
Support from the Deutsche Forschungs\-ge\-mein\-schaft (DFG), the Fonds der
Chemischen Industrie and the Ministry of Science, Research and the Arts of
Baden-W\"urttemberg (AZ: 24-7532.23-11-11/1) is gratefully acknowledged.

\end{document}